\documentclass[epj]{svjour}
\usepackage{graphicx}
\usepackage{amsfonts}
\begin{document}
\title{Self-organized criticality in nonconservative mean-field sandpiles}
\titlerunning{SOC in nonconservative sandpiles}
\authorrunning{D.E. Juanico}
\author{Dranreb Earl Juanico
}                     
%
%
\institute{Rm 3111 Complex Systems Theory Group, National Institute
of Physics, \\University of the Philippines, Diliman, Quezon City,
Philippines 1101\\\email{djuanico@gmail.com}}
\date{Received: date / Revised version: date}
%
\abstract{ A mean-field sandpile model that exhibits self-organized
criticality (SOC) despite violation of the grain-transfer
conservation law during avalanches is proposed. The sandpile
consists of $N$ agents and possesses background activity with
intensity $\eta\in[0,1]$. Background activity is characterized by
transitions between two stable agent states. Analysis employing
theories of branching processes and fixed points reveals a
transition from sub-critical to SOC phase that is determined by
$\eta N$. The model is used to explain the school size distribution
of free-swimming tuna as a result of population depletion.
} 
\maketitle
\section{Introduction}\label{intro}
Prototypical SOC sandpile models have attracted several researchers
due to their inherent simplicity yet non-trivial behavior. Dynamical
mean-field theory has been proven to offer the most effective
elucidation of how SOC works~\cite{1}. One of the key ingredients
found to bring about SOC in sandpiles is grain-transfer
conservation. Tsuchiya and Katori offered rigorous proof that
violation of the conservation law frustrates the criticality of
Abelian sandpiles~\cite{2}. A mean-field treatment of the Manna
sandpile known as the self-organized branching process (SOBP) has
been proposed, further demonstrating that nonconservation disrupts
criticality~\cite{3}. Breaking of SOC due to nonconservation has
also been later demonstrated numerically and analytically in the
Olami-Feder-Christensen earthquake model~\cite{4}. The forest-fire
model (FFM), which does not have conservation laws, was also
proposed as a model for SOC~\cite{5}. However, it was ultimately
proven, via analysis of Lyapunov exponents~\cite{6} and later by
means of renormalization group approach~\cite{7}, that FFM does not
exhibit SOC.

Signatures of scale invariance and criticality in biology, ecology,
and a wide range of other animate complex systems have been linked
to the principles behind power-law generating SOC models such as the
sandpile model~\cite{8}. Practically, biological systems are not
expected to be conservative either because they are constantly
interacting with the environment and are far from equilibrium. These
systems are therefore not supposed to display SOC.

Here, I propose a mean-field sandpile model that displays
criticality despite violation of grain-transfer conservation. This
model takes off from the conceptual difficulty with typical sandpile
models in defining fluctuations between stable states because
passive sites only switch states when an avalanche crosses them or
when a grain is sprinkled on them. Thus, sites in this model are
turned into \emph{active agents}. Demonstrating criticality in
self-organizing systems that are nonconservative enhances the
synthetical capacity of SOC theory in the arena of biocomplexity.

Section~\ref{sec:model} describes the nonconservative model.
Section~\ref{sec:results} presents and discusses the results of
analysis and simulations of the model.
Section~\ref{sec:recommendations} elaborates salient assumptions and
possible areas for further analysis and extension. Lastly,
Section~\ref{sec:summary} summarizes and concludes the paper.
\section{Model} \label{sec:model}
\subsection{Framework}
Inspired by biocomplexity, a sand grain assumes the role of a
stimulus---for instance an environmental trigger---that is stored,
integrated, and transferred from one agent to another. The build-up
of stored stimuli induces agent response, which activates the
transfer of stimulus through neighboring agents. Stimulus
propagation corresponds to avalanches.

In accordance with Manna sandpile rules and studies of excitable
media~\cite{9}, each agent in the system may be in any three states
$\{\rm{refractory}, \rm{quiescent}, \rm{excited}\}$ mapped to
$\{z\}=\{0,1,2\}$, as follows:
\begin{itemize}
  \item[$\bullet$] $ {\rm quiescent} \mapsto z=1$
  \item[$\bullet$] $ {\rm excited} \mapsto z=2$
  \item[$\bullet$] $ {\rm refractory} \mapsto z=0$
\end{itemize}

By storage and integration of stimulus, refractory agents may turn
quiescent and quiescent agents may become excited. If threshold
$z_{\rm{th}}=1$, then an excited agent's stimulus level $z >
z_{\rm{th}}$, so that it fires a response in order to relax back to
the stable states $z=0, 1$. When this happens, an avalanche ensues.
``Avalanche" is here interpreted as \emph{clustering} in the sense
that the agents excited during an avalanche are all ``behaviorally
matched" to the initial agent that triggers the avalanche.

This model differs from typical sandpile models because it allows
for the possibility that, apart from excitations triggered along the
path of the avalanche, a refractory agent can spontaneously become
quiescent and vice versa, albeit at a much slower pace than the
avalanche itself. This independent but slow process is what is
referred hereafter as \emph{background activity}.

The system is a population of $N=2^{n+1}-1$ excitable agents, where
$n$ corresponds to the maximum number of generations of excitation
allowed during a single avalanche, so that $n$ is a boundary
condition. The model has no explicit dependence on any spatial
lattice configurations, which makes analysis straightforward due to
the absence of nearest-neighbor correlations and is a more accurate
representation of socially interacting systems made up of entities
that are constantly in motion such as animal groups. The mean-field
nature of the model also does not require a physical boundary. Fish
schools, for instance, traverse a seemingly limitless oceanic space.
\subsection{Mean-Field Transition Rules}
The population dynamically reconfigures the landscape of agent
states when there is no avalanche. A quiescent agent turns
refractory with probability $\lambda$, whereas a refractory agent
turns quiescent with probability $\eta$. These transitions between
refractory and quiescent states (stable agent states) is the
background activity.

The time-dependent density $q(t)$ of quiescent agents also
corresponds to the probability that an external stimulus injected at
time $t$ into a randomly selected agent initiates excitation,
because only quiescent agents can become excited at any given time.
If at time $t$ an excited agent indeed emerges, time is frozen and
avalanche ensues. Freezing time follows from the assumption that
background activity takes place much slower than an
avalanche---known as ``timescale separation," which is a typical
assumption in sandpile models~\cite{1}.

During an avalanche, an excited agent stimulates at most two other
agents. With probability $\alpha$ the excited agent turns refractory
by transferring $2$ units of stimuli to two randomly chosen
neighbors. With probability $\beta$ the excited agent turns
quiescent by dissipating only $1$ unit of stimulus to one random
neighbor. Incorporating a probability $\epsilon=1-\alpha-\beta$ that
an excited agent turns refractory by inwardly absorbing $2$ units of
stimuli without subsequently stimulating other agents, the transfer
rule becomes nonconservative. Conservation law is therefore violated
for $\epsilon > 0$.

The nonconservative transfer rule repeats until all excited agents
are depleted, after which the frozen time takes off again and the
population builds up for the next avalanche. There is a limit $n$ to
the number of generations of transfers during an avalanche. At the
$n$-th generation, any remaining excited agents mandatorily absorbs
stimuli without further stimulation of other agents. All the
aforementioned processes are summarized in
table~\ref{tab:transition_rules}.
\begin{table}
  \caption{Mean-field transition rules of the nonconservative branching
  model with corresponding transition probabilities. The central digits
  for the avalanche rules correspond to the excited agent and the flanking
  digits are the two random neighbors, in no particular order. The sum of
  transition probabilities for the avalanche and nonconservative rules
  is $q$, which is, self-consistently, the probability that the avalanche
  had initiated.}\label{tab:transition_rules}
  \centering
  \begin{tabular}{lll}
    \hline
    Process & Rule & Transition probability \\\hline
    Background  & $1\rightarrow 0$ & $\lambda q$ \\
      & $0\rightarrow 1$ & $\eta(1-q)$ \\
    Avalanche & $121\rightarrow 202$ & $\alpha q^3$ \\
     (branching process) & $021\rightarrow 102$ & $2\alpha q^2(1-q)$ \\
     & $020\rightarrow 101$ & $\alpha q(1-q)^2$ \\
     & $121\rightarrow 112$ & $\beta q^3$ \\
     & $021\rightarrow 012$ & $2\beta q^2(1-q)$ \\
     & $020\rightarrow 011$ & $\beta q(1-q)^2$ \\
    Nonconservation & $2\rightarrow 0$ & $\epsilon q=(1-\alpha-\beta)q$ \\
    \hline
  \end{tabular}
\end{table}
\subsection{Dynamics and Branching Process}
Given the transition rules, the density $q$, assumed to be a
continuous dynamical variable, satisfies the stochastic dynamical
equation
\begin{equation}\label{eq:quiescent_derivative}
    \frac{\mathrm{d}q}{\mathrm{d}t}=(1-q)\eta-q\lambda+\mathcal{A}(q;\alpha,\beta)+\xi(t)/N .
\end{equation}
The noise term $\xi/N$ arises from the stochastic transition rules,
accounting for fluctuations around mean values assumed to hold in
the mean-field calculations. It appropriately vanishes in the
large-$N$ limit. The term $\mathcal{A}(q; \alpha,\beta)$ represents
the change in $q(t)$ due to redistribution of stimulus by an
avalanche at time $t$. Treating an avalanche as a branching process,
and following closely the analysis in~\cite{3},
$\mathcal{A}(q;\alpha,\beta)$ satisfies
\begin{equation}\label{eq:avalanche_term}
    N\mathcal{A}= 1-\sigma^n - \frac{\epsilon
    q}{1-(1-\epsilon)q}\left[1+\frac{1-\sigma^{n+1}}{1-\sigma}-2\sigma^n\right],
\end{equation}
where $\sigma$ is the branching parameter derived from the following
definition
\begin{equation}\label{eq:def_branching}
    \sigma = \sum_k k\pi_k ,
\end{equation}
wherein from table~\ref{tab:transition_rules}, the branching
probability $\pi_k$ that an excited agent subsequently stimulates
$k$ other agents is
\begin{equation}\label{eq:def_branch_prob}
    \pi_k = \alpha q\delta_{k,2}+\beta
    q\delta_{k,1}+\left[1-(1-\epsilon)q\right]\delta_{k,0} ,
\end{equation}
with $\delta_{i,j}$ being the Kronecker delta. The first and second
terms in eq.~(\ref{eq:def_branch_prob}) are the sum of transition
probabilities of the first three avalanche rules and of the last
three avalanche rules listed in table~\ref{tab:transition_rules},
respectively. The last term is the total probability coming from the
nonconservation rule and the case wherein a subsequently stimulated
agent is refractory, which both result to $k=0$. Substitution of
eq.~(\ref{eq:def_branch_prob}) into eq.~(\ref{eq:def_branching})
gives
\begin{equation}\label{eq:sigma}
    \sigma = (2\alpha+\beta) q ,
\end{equation}
which depends on $\alpha$ and $\beta$, and is proportional to
density $q$. In this model it is assumed that $\alpha$ and $\beta$
remain fixed throughout so that the only dynamic variable is $q$.
Consequently,
\begin{equation}\label{eq:dsigma_dt}
    \frac{\mathrm{d}\sigma}{\mathrm{d}t} =
    (2\alpha+\beta)\frac{\mathrm{d}q}{\mathrm{d}t}.
\end{equation}

A branching process is sub-critical when $\sigma<1$, and
consequently avalanches have sizes always smaller than a finite
cutoff size. On the other hand, a branching process is
supra-critical when $\sigma>1$, and consequently avalanches with
sizes as large as the system itself are formed almost with
certainty. Hence, $\sigma=1$ is a critical value that results to a
critical branching process~\cite{10}. The avalanche size
distribution emerging from a critical branching process is expected
to be power law. Since $\sigma$ only varies when $q$ changes, the
$q$ determines whether the branching process is sub-critical,
critical, or supra-critical.

To keep the model minimal, the number of parameters in
eq.~(\ref{eq:quiescent_derivative}) is reduced by introducing the
relation
\begin{eqnarray}\label{eq:matching_condition}\nonumber
    \frac{\lambda}{\eta} &=& \sigma/q -1\\
    &=& 2\alpha+\beta - 1\, ,
\end{eqnarray}
which effectively couples the background activity to the avalanche
(branching process). Eq.~(\ref{eq:matching_condition}) is a
phenomenological assumption which essentially guarantees that $q(t)$
in the steady state approaches a value that makes $\sigma = 1$. This
is an indication that the model self-organizes to its critical
state. In order to analytically prove that the model indeed exhibits
SOC, the critical state must be an attractive fixed point in phase
space~\cite{7}. The phase portrait is simply a plot of $\dot{q} :=
\rm{d}q/\rm{d}t$ versus $q$. Since $\sigma\propto q$, then the phase
portrait can equivalently be visualized by plotting
$\dot{\sigma}:=\rm{d}\sigma/\rm{d}t$, defined in
eq.~(\ref{eq:dsigma_dt}), versus $\sigma$.
\subsection{Avalanche Size Distribution}
\begin{figure}
  \centering
  \includegraphics[width=\columnwidth]{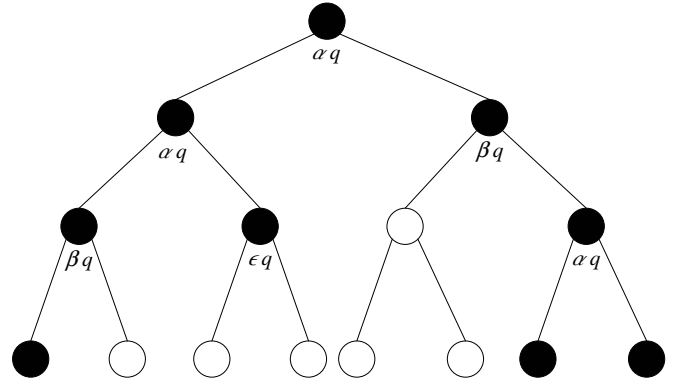}\\
  \caption{Avalanche for $n=3$ shown as a branching tree. Shaded circles correspond to
  excited agents. The avalanche size $s=9$ corresponds to the total number of shaded
  circles.}\label{fig:branching_tree}
\end{figure}

An avalanche resulting from the rules of the model is illustrated as
a branching tree in figure~\ref{fig:branching_tree}. The initial
excited agent is the topmost node, giving rise to two subsequently
excited nodes with probability $\alpha q$. One of these excited
nodes in turn generates one excited node with probability $\beta q$,
while the other generates two excited nodes with probability $\alpha
q$. One of such excited nodes do not generate a subsequently excited
node with probability $\epsilon q$ due to nonconservation. Setting
$n=3$, all excited nodes at three levels below the topmost node do
not further generate excited nodes. In the example shown, the
avalanche size $s=9$.

\begin{figure}
  \centering
  \includegraphics[width=\columnwidth]{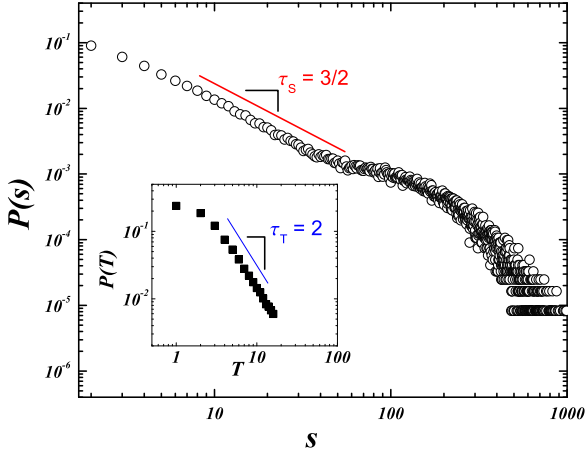}\\
  \caption{Avalanche size distribution of a nonconservative system with
  $\epsilon=0.2$ and $\eta N\approx 8192$. A power-law with exponent $\tau_{\mathrm S}=3/2$
  is drawn as guide to the eye. The exponential fat tail is an artifact
  of the finiteness of the system. The inset graph is the distribution of
  avalanche lifetimes, which is the number of excitation generations before
  the avalanche dies out. A power law with exponent $\tau_{\mathrm T}=2$ is
  drawn as guide.}\label{fig:simulated_dist}
\end{figure}

The branching probability $\pi_k$ defined in
eq.~(\ref{eq:def_branch_prob}) essentially describes the likelihood
that an excited agent subsequently generates $k\in\{0,1,2\}$ excited
agents in the succeeding generation. Using $\pi_k$ the avalanche
size distribution $P(s)$ is calculated via a generating functional
formalism. A generating function $\frak{F}_m(\omega)$ for $P(s)$
after $m$ generations is defined as
\begin{equation}\label{eq:genf_m}
    \frak{F}_{m}(\omega) = \sum_{s=1}^{\infty} P(s)\omega^s\, .
\end{equation}
Incidentally, $\frak{F}_m$ is also the $m$-th iterate of the
generating function $\frak{F}_1 := \frak{F}$, defined as
\begin{equation}\label{eq:genf_1}
    \frak{F}(\omega) = \sum_{s=1}^{\infty}\pi_{s-1}\omega^s = \omega\sum_{s=1}^{\infty}\pi_{s-1}\omega^{s-1}.
\end{equation}
Also by definition, $\frak{F}_{m+1} = \frak{F}(\frak{F}_m)$, such
that from eq.~(\ref{eq:genf_1}) one obtains
\begin{eqnarray}\nonumber
    \frak{F}_{m+1}(\omega) = \omega\sum_{s=1}^{\infty}
    \pi_{s-1}\left[\frak{F}_m(\omega)\right]^{s-1},
\end{eqnarray}
which simplifies to
\begin{equation}\label{eq:recursion}
    \frak{F}_{m+1}(\omega) = \omega\left\{\alpha q\frak{F}_m^2+\beta
    q\frak{F}_m+\left[1-(1-\epsilon)q\right]\right\},
\end{equation}
following from the definition of $\pi_k$ in
eq.~(\ref{eq:def_branch_prob}). For large enough $m$, the theory of
branching processes asserts a self-consistency relation wherein
$\frak{F}_{m+1}\simeq\frak{F}_m$, so that solving for $\frak{F}_m$
in eq.~(\ref{eq:recursion}) yields
\begin{equation}\label{eq:genf_iterate}
    \frak{F}_m(\omega) =
    \frac{1-b\omega-\sqrt{1-2b\omega+a\omega^2}}{2\alpha q\omega },
\end{equation}
where $a=\beta^2q^2-4\alpha q[1-(1-\epsilon)q]$, and $b=\beta q$.
Binomial expansion of eq.~(\ref{eq:genf_iterate}) around its
singularity $\omega=0$, one obtains a power series similar to
eq.~(\ref{eq:genf_m}). The coefficients of this expansion correspond
to $P(s)$ from eq.~(\ref{eq:genf_m}). In a more compact form the
solution can be expressed as a recurrence relation
\begin{equation}\label{eq:P_s}
    P(s) = \frac{1}{s+1}\left[(2s-1)bP(s-1)-(s-2)aP(s-2)\right].
\end{equation}
Eq.~(\ref{eq:P_s}) may be shown, by graphical inspection, to have
the asymptotic behavior $P(s)\sim s^{-3/2}$ for $s\gg 1$ if $a =
2b-1$; a condition that can equivalently be expressed as
\begin{eqnarray}\label{eq:critical_condition}\nonumber
    Q(q) &=& a+2b-1\\
    &=& \beta^2 q^2-4\alpha q[1-(1-\epsilon)q]-2\beta q
    +1\\\nonumber
    &=& 0\, .
\end{eqnarray}
Numerical simulation of the model also confirms the asymptotic
behavior of eq.~(\ref{eq:P_s}), as shown in
figure~\ref{fig:simulated_dist}, for degree of nonconservation
$\epsilon=0.2$ and $\eta N\approx 8192$.

The function $Q(q)$ is parabolic in terms of $q$ and has a unique
root at $q_{\rm c}=(2\alpha+\beta)^{-1}$, which makes $\sigma=1$.
Hence, eq.~(\ref{eq:critical_condition}) is a criticality condition.
Furthermore, since $\sigma\propto q$, then $Q(q)$ can alternatively
be expressed as $Q(\sigma)$ which has a root at $\sigma=1$.
\section{Results and Discussion}\label{sec:results}
The stationary behavior of the model is examined using a
well-established geometrical theory of fixed points~\cite{11}.
Nonlinear differential equations such as eq.~(\ref{eq:dsigma_dt})
may be analyzed graphically in terms of vector fields. In this
framework $\dot{\sigma}$ is interpreted as a ``velocity vector" at
each possible $\sigma$ value. Plotting $\dot{\sigma}$ versus
$\sigma$ is the phase portrait of the model. The fixed point
$\sigma^*$ is the value of $\sigma$ at which $\dot{\sigma}=0$. The
trajectory of the vector around the neighborhood of this fixed point
is directed to the right where $\dot{\sigma}>0$, and to the left
where $\dot{\sigma}<0$. This means that if $\dot{\sigma}$ is
increasing around the neighborhood of $\sigma^*$, then the fixed
point is repulsive. On the other hand, if $\dot{\sigma}$ is
decreasing, then the fixed point is attractive. Linearization of
eq.~(\ref{eq:dsigma_dt}) actually yields
\begin{eqnarray}\nonumber
    \lim_{\sigma\rightarrow\sigma^*}\frac{\partial}{\partial\sigma}\frac{\mathrm{d}\sigma}{\mathrm{d}t}<0\,
    .
\end{eqnarray}

\begin{figure}
  \centering
  \includegraphics[width=\columnwidth]{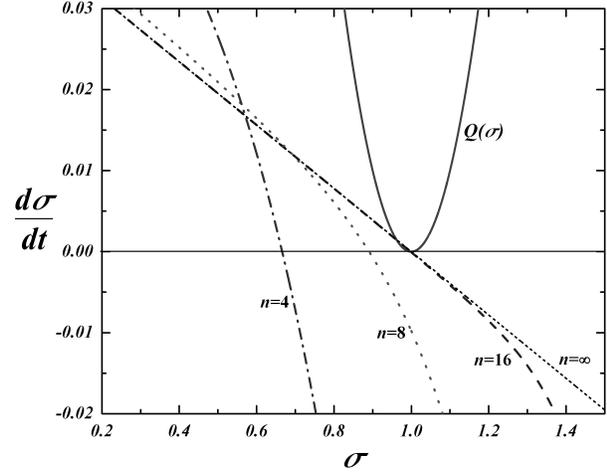}\\
  \caption{Phase portrait showing $\dot{\sigma}$ versus $\sigma$,
  of a nonconservative system with $\epsilon=0.25$ and $\eta=0.03125$ for
  different $n$: 4 (chain curve), 8 (dotted curve), 16 (dashed curve), and
  $\infty$ (broken line). The root of $\dot{\sigma}$ corresponds to the fixed point
  value $\sigma^*$ of the branching parameter. The solid parabolic curve
  is the function $Q(\sigma)$ with root at $\sigma=1$. For $n\geq 16$,
  $\sigma^*=1$, which indicates that the nonconservative system evolves
  towards its critical state via self-organization.}\label{fig:dsigma-dt_sigma}
\end{figure}

Shown in figure~\ref{fig:dsigma-dt_sigma} are the phase portraits of
the model for different $n$, which are monotonically decreasing.
Thus, the fixed points of eq.~(\ref{eq:dsigma_dt}) are indeed
dynamically attractive. Therefore, apart from fluctuations, the
model spontaneously approaches the state defined by its fixed
point---the very idea of self-organization.

Also illustrated in figure~\ref{fig:dsigma-dt_sigma} is the function
$Q(\sigma)$ with a root at $\sigma=1$. For $n=4$ and $n=8$,
obviously $\sigma^*<1$, implying that the system self-organizes
towards a sub-critical state. However, from $n=16$ up to
$n\rightarrow\infty$, $|~\sigma^*~-~1~|\ll 1$. A nonconservative
system achieves criticality even if its size is finite. The rapid
approach of $\sigma^*$ towards $1$ from $n=8$ to $n=16$ and the
decelerating change in $\sigma^*$ from $n=16$ to
$n\rightarrow\infty$ indicates a phase transition.
\begin{figure}
  \centering
  \includegraphics[width=\columnwidth]{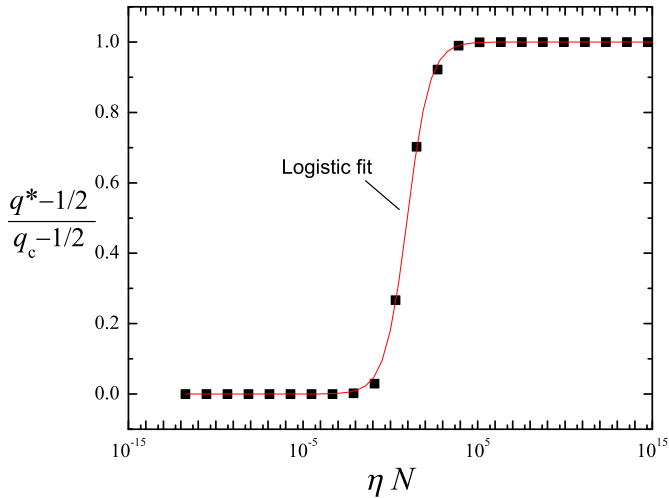}\\
  \caption{Phase transition in the nonconservative branching model. Data
  points result from numerical analysis of the fixed points $q^*$ of
  Equation~\ref{eq:quiescent_derivative} neglecting fluctuations $\xi/N$,
  for various $\epsilon\in(0,0.5)$, $\eta\in(0,1]$, and system size $N$.
  The curve is a logistic fit of the form
  $y(x)=1-[1+(x/\theta)^{\gamma}]^{-1}$ where $y=(q^*-1/2)/(q_{\mathrm c}-1/2)$
  and $x=\eta N$.}\label{fig:phase_transition}
\end{figure}

Indeed, as shown in figure~\ref{fig:phase_transition}, a phase
transition of the model exists. The data points are derived from
numerical fixed-point analysis of
eq.~(\ref{eq:quiescent_derivative}), assuming that the noise term
$\xi/N$ is very small to be significant. For a wide range of
parameter values: $\epsilon\in(0,1/2)$ and $\eta\in(0,1]$, a
transition surfaces out from the relation between  $\eta N$ and
$(q^*\!-\!1/2)/(q_{\rm c}\!-\!1/2)$. The resultant plot is fitted by
a logistic curve
\begin{equation}\label{eq:logistic_fit}
    \frac{q^*-1/2}{q_{\mathrm c}-1/2} = 1-\left[1+\left(\frac{\eta
    N}{\theta}\right)^{\gamma}\right]^{-1} ,
\end{equation}
where $\theta=9.31\pm 0.27$ and $\gamma=0.68\pm 0.01$ (goodness of
fit: $\chi^2/\rm{DoF}=4\times10^{-5}$, $R^2=0.99986$, no weighting).
Eq.~(\ref{eq:logistic_fit}) suitably describes two limiting cases.
The first one is $\eta\rightarrow 0$, which results to
$q^*\rightarrow 1/2$. This limiting case is equivalent to having no
background activity, as in the SOBP model by Lauritsen, Zapperi \&
Stanley~\cite{3}. They have similarly found that the steady-state
(fixed-point) value is $q^*=1/2$, irrespective of any degree of
nonconservation $\epsilon$. Consequently, their sandpile model is
always sub-critical whenever $\epsilon >0$.

The second limiting case is $nN\gg\theta$, which leads to
$q^*\rightarrow q_{\rm c}$. At this limit, the nonconservative
system is always critical. A closer look at
figure~\ref{fig:phase_transition} reveals that the data points lie
considerably close to $1$ starting at $\eta N\sim 10^4$. This means
that if the population is very large, say $N=10^9$, then for a wide
range of $\eta\in(10^{-5},1]$ the nonconservative system achieves
SOC. Simulation of the model for various degrees of nonconservation
$\epsilon$ yields an avalanche size distribution of the form plotted
in figure~\ref{fig:simulated_dist}.

\begin{figure}
  \centering
  \includegraphics[width=\columnwidth]{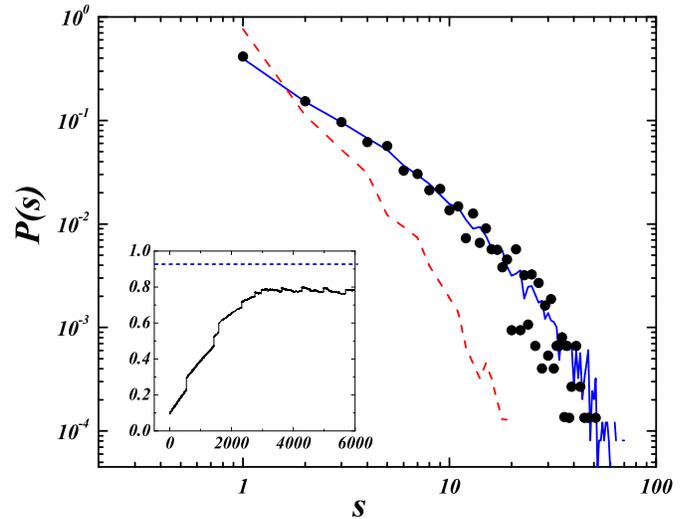}\\
  \caption{School size distribution of free-swimming tuna (shaded circles),
  and of a nonconservative system with $\epsilon=0.15$ and $\eta N=409.5$
  taken after $2^{15}$ time steps (solid curve). Also shown is a size
  distribution resulting from an equivalent SOBP model (dashed curve) having
  the same value of $q_{\mathrm c}=0.952$ as the nonconservative model.
  Inset graph reveals the evolution of $q(t)$. Its steady-state value $q^*$ is
  less than $q_{\mathrm c}$ (dashed line), implying that the nonconservative
  system is sub-critical.}\label{fig:tuna}
\end{figure}

In the context of ecological systems, large $N$ may easily be
satisfied by clumps of microorganisms such as cellular aggregates in
culture~\cite{12}. However, in macroscopic systems such as animal
groups, this may not be the case because of several ecological
factors, such as the presence of predators and destruction of
habitats, which curtail the proliferation of a certain species. A
good example is a population of tuna fishes studied by Bonabeau et
al.~\cite{13}. The school size distribution $P(s)$ found from this
study is plotted in fig.~\ref{fig:tuna}, fitted by the
nonconservative model with $\epsilon=0.15$ and $\eta N=409.5$. This
$\eta N$ value maps to a sub-critical phase on the phase diagram
shown in figure~\ref{fig:phase_transition}, explaining the prominent
exponential tail in $P(s)$. Also shown is a curve from an equivalent
SOBP model~\cite{3}, exhibiting less agreement with the data.

A school size distribution having an exponential tail implies that
schools cannot be expected to be larger than a finite cutoff size.
However, large schools or groups of animals are required for
sustainability and wildlife preservation as it is believed that
group size affects reproductive success of animals, most especially
endangered species.

Lastly, it is also interesting to note that for $\eta
N\simeq\theta$, the $q^*$ value is highly uncertain to small
variations in either $\eta$ or $N$. This is the regime describing
the abrupt rise from 0 to 1 in figure~\ref{fig:phase_transition}.
\section{Recommendations}\label{sec:recommendations}
A key assumption in the model rests upon having the branching
parameter defined in eq.~(\ref{eq:sigma}) sufficiently close to its
critical value of 1. With $\alpha$ and $\beta$ fixed, $\sigma=1$ is
obtained solely by the incorporation of a background activity. This
background activity is coupled to the avalanche via a matching
condition [eq.~(\ref{eq:matching_condition})] to guarantee that
$q=(2\alpha+\beta)^{-1}$ at the steady state. As a modification, one
could assume instead that both $\alpha$ and $\beta$ also change with
$t$, perhaps due to some local feedback mechanism that adjusts the
branching probability depending on the history of an agent's
activity. Hence in general eq.~(\ref{eq:dsigma_dt}) may be written
as
\begin{eqnarray}\nonumber
\frac{{\rm d}\sigma}{{\rm d}t} =
\left(2\frac{\partial\alpha}{\partial
t}+\frac{\partial\beta}{\partial t}\right)q +
(2\alpha+\beta)\frac{\partial q}{\partial t}\, .
\end{eqnarray}
The implementation of the above modification lies in defining the
feedback mechanism that gives $\partial\alpha/\partial t$ and
$\partial\beta/\partial t$. Lastly, the model can be extended to the
case wherein $\{z\}=\{0,1,2,\ldots,z_{\rm{th}},z_{\rm{th}}+1\}$ and
$z_{\rm{th}}>1$. Preliminary research is underway to evaluate the
performance of the model with these modifications.
\section{Summary and Conclusion}\label{sec:summary}
In summary, I have proposed a mean-field sandpile model coupled with
a background activity that is characterized by local fluctuations
between stable agent states. Even in the presence of violation of
the grain-transfer rule, the sandpile can self-organize to a
critical state. This result addresses the long-standing issue of
whether SOC is possible in nonconservative sandpiles. The model is
applied to explaining the truncated school size distribution
observed for free-swimming tuna and provides insight into the
effects of population depletion on the aggregation capacity of
clustering animals. Lastly, some recommendations for extending the
model have been rendered.

\begin{acknowledgement}
The author wishes to acknowledge the OVCRD in UP Diliman for Grant
No. 050501 DNSE, and the PCASTRD-DOST for funding this research.
\end{acknowledgement}

\end{document}